\documentclass[aps, pre, onecolumn, superscriptaddress, tightenlines, longbibliography]{revtex4-2}

\usepackage{amsmath}
\usepackage{amssymb}
\usepackage{mathtools}
\usepackage{physics}
\usepackage{bm}
\usepackage{threeparttable}
\usepackage{graphicx}
\usepackage{booktabs}
\usepackage{multirow}
\usepackage{epstopdf}
\usepackage{natbib}
\usepackage[colorlinks=true,linkcolor=blue,citecolor=blue,urlcolor=blue]{hyperref}

\begin{document}

    \title{Non-Monotonic Exciton Transport Efficiency in Multi-layer Ring-shaped Quantum Networks inspired by Photosystem I}

    \author{Yi Li}
    \affiliation{Department of Physics, School of Physics and Astronomy, Beijing Normal University, and Key Laboratory of Multiscale Spin Physics(Beijing Normal University), Ministry of Education, Beijing 100875, China}

    \author{Rui Li}
    \affiliation{Department of Physics, School of Physics and Astronomy, Beijing Normal University, and Key Laboratory of Multiscale Spin Physics(Beijing Normal University), Ministry of Education, Beijing 100875, China}

    \author{Kai-Ya Zhang}
    \affiliation{Department of Physics, School of Physics and Astronomy, Beijing Normal University, and Key Laboratory of Multiscale Spin Physics(Beijing Normal University), Ministry of Education, Beijing 100875, China}

    \author{Ming-Zi Jin}
    \affiliation{Department of Physics, School of Physics and Astronomy, Beijing Normal University, and Key Laboratory of Multiscale Spin Physics(Beijing Normal University), Ministry of Education, Beijing 100875, China}

    \author{Tao Liu}
    \affiliation{Department of Applied Physics, School of Physics and Optoelectronics, South China University of Technology, Guangzhou 510640, China}

    \author{Qing Ai}
    \affiliation{Department of Physics, School of Physics and Astronomy, Beijing Normal University, and Key Laboratory of Multiscale Spin Physics(Beijing Normal University), Ministry of Education, Beijing 100875, China}
    \email{aiqing@bnu.edu.cn}
    \begin{abstract}
    Photosystem I (PSI) is one of the most efficient natural photosynthetic complexes and serves as a structural prototype for understanding excitation energy transfer (EET) and designing artificial light-harvesting systems. 
    Here, we construct a multilayer elliptical-conical quantum network inspired by PSI to investigate how antenna architecture affects EET.  
    Excitation dynamics are simulated using the quantum jump approach combined with coherent modified Redfield theory (QJA-CMRT). 
    The transfer efficiency exhibits a pronounced nonmonotonic dependence on pigment density, reaching a maximum of 0.86 at $x=6$ in the five-layer network. The optimal configuration contains 91 pigments, close to the number of core antenna pigments in natural PSI. 
    Our combined density‑of‑states and kinetic analyses reveal that this nonmonotonic dependence originates from a trade‑off: at low pigment densities, enhanced band overlap promotes energy transfer, whereas at high densities, excessive exciton delocalization compromises funnel directionality and reduces efficiency; the optimal configuration is consequently achieved at $x=6$.
    These findings provide insights into PSI antenna organization and offer guidelines for designing efficient artificial light-harvesting systems.
    \end{abstract}


\maketitle

\section{Introduction}
Photosynthesis is the main biological process that captures solar energy and converts it into chemically available energy, which maintains most life forms on earth~\cite{Blankenship2014}. The initial step of photosynthesis is excitation energy transfer (EET), in which light absorbed by antenna pigments is transferred to the reaction center where charge separation occurs~\cite{vangrondelle2006,mirkovic2017,song2021excitonic,nguyen2023charge}. The natural light harvesting system achieves high efficiency in noisy and fluctuating biological environments, making it a valuable model for basic biophysics and artificial energy conversion design~\cite{Lambert2013nphys,Blankenship2014,mirkovic2017,romero2017quantum,Cao2020sciadv}. In natural photosynthetic complexes, photosystem I (PSI) is particularly important because it combines a large antenna pigment network with fast excitation transmission and efficient reaction center capture~\cite{Croce2013,jordan2001,nelson2015,akhtar2021}. Therefore, understanding how the PSI antenna structure controls EET provides a direct way to identify the structural principles~\cite{huchzermeyer1986direct,huchzermeyer1981coupling,huchzermeyer1980bicarbonate} that can be migrated to artificial photosynthesis and molecular excitation subsystems.

From an evolutionary perspective, while the PSI core antenna is remarkably conserved across oxygenic phototrophs, the peripheral light-harvesting system exhibits substantial diversity reflecting adaptation to distinct photic niches~\cite{Tian2024ijms}. In cyanobacteria, the ancestral oxygenic phototrophs, PSI exists as a trimer with $\sim$96 chlorophylls per monomer and no additional membrane-integral antennae~\cite{karapetyan1999,jordan2001}. In contrast, land plants have evolved four LHCI subunits surrounding the core, increasing the pigment count to $\sim$170--200 chlorophylls to balance light harvesting with linear electron flow requirements~\cite{Zhang2024natplants,Naschberger2024natplants}. Green algae display intermediate plasticity with 6--10 LHCIs, whereas red-lineage algae, including diatoms and coccolithophores, exhibit extreme antenna expansions~\cite{kato2024elife,Feng2024jipb}. Recent cryo-EM structures reveal that the coccolithophore \textit{Emiliania huxleyi} possesses a massive PSI-fucoxanthin chlorophyll \textit{a/c}-binding protein (FCPI) supercomplex with 35--38 LHCIs and over 800 pigments~\cite{shen2025science,sun2025sciadv}. Similarly, seagrasses like \textit{Posidonia oceanica} have genetically adapted their PSI antenna size and spectral properties to thrive in dim, spectrally shifted seabed light~\cite{Amelii2026natcomm}. Despite this four- to five-fold expansion in light-harvesting cross-section relative to land plants, these supercomplexes maintain a quantum efficiency of approximately 95\%~\cite{shen2025science,sun2025sciadv,Akhtar2025cellrep}.

The high efficiency of PSI is closely related to its pigment arrangement, site energy distribution and electron coupling of protein regulation~\cite{Lee2007science,wang2024incoherent}. Structural studies have shown that the PSI core contains a dense chlorophyll antenna around the reaction center, and its pigment position and orientation are constrained by the protein skeleton~\cite{jordan2001,nelson2015}. More recently, the cryo-EM structures of coccolithophore PSI have revealed the largest PSI-antenna supercomplexes known to date, with 35-38 peripheral LHC antennae organized in multiple concentric layers, containing over 800 pigments and achieving a quantum efficiency of approximately 95\%~\cite{shen2025science,sun2025sciadv}. Spectroscopic and theoretical studies further show that the excitation transmission in PSI is not controlled by a single path~\cite{song2019vibronic}, but is determined by a complex energy landscape, which is jointly shaped by exciton delocalization, low-energy chlorophyll state, environmental relaxation and reaction center capture~\cite{jordan2001,akhtar2021,zhu2024quantum}. These characteristics indicate that PSI is not just a collection of independent pigments. On the contrary, it functions as a structured exciton network, in which geometric structure and energy jointly determine the transmission efficiency. Therefore, the core issue is how changes in pigment density and spatial organization affect the balance between beneficial exciton coupling and non-productive losses.
The high efficiency is physiologically remarkable given the extremely high in vivo pigment concentration ($\sim$0.5~M) within the protein scaffold~\cite{Croce2013}. In solution, such concentrations would cause severe concentration quenching, reducing excited-state lifetimes from $\sim$6~ns to $<$100~ps~\cite{Croce2013}. However, the evolutionarily conserved protein matrix precisely controls inter-pigment distances and orientations, suppressing non-radiative decay while facilitating fast energy transfer~\cite{nelson2015,Naschberger2024natplants}. Understanding this structural principle is crucial for elucidating the evolutionary optimization of photosynthetic efficiency and for designing bio-inspired artificial light-harvesting systems~\cite{huchzermeyer1986direct,huchzermeyer1981coupling,huchzermeyer1980bicarbonate}.

A variety of theoretical models have been developed to describe EET in photosynthesis systems. The F{\"o}rster resonant energy transfer theory provides a simple and physically clear description~\cite{papenbrock2016current} in the weak coupling and incoherent hopping regions, where the transfer rate is determined by electron coupling and donor-acceptor spectral overlap~\cite{vanAmerongen2000,may2011,vangrondelle2006}. The Redfield-type and modified Redfield methods are more applicable when exciton delocalization and system-environment interactions need to be more clearly addressed~\cite{may2011,renger2009,ishizaki2009,hwang-fu2015}. The hierarchical method of the equation of motion can provide a high-precision description of quantum dissipation dynamics, but for large pigment networks, the computational cost is considerable~\cite{ishizaki2009}. The environment-assisted quantum transport model also shows that moderate noise or dephasing can enhance transmission by suppressing localization and helping to stimulate escape from inefficient paths~\cite{plenio2008,mohseni2008,chin2013,Maly2019chemphys,dong2012coherent}. These methods greatly advance the understanding of photosynthetic EET, but the direct structural interpretation of how antenna density, exciton bandwidth, density of states overlap, and fluorescence loss together determine the optimal size of PSI-like networks remains to be established. 

Previous studies have emphasized that pigment aggregation can improve transmission efficiency by enhancing electron coupling, promoting exciton delocalization, and creating resonant energy transfer channels~\cite{cao2009,ai2013clustered,mattioni2021}. However, aggregation is not always beneficial. When the pigment is too dense to accumulate, the exciton energy band may be excessively broadened, the energy direction of the energy funnel may be weakened, and the excitation may stay longer in the area where the transmission is not active or easy to lose. In real photosynthesis antennas, changes in pigment accumulation, protein conformation, and aggregation state also promote non-photochemical energy dissipation or fluorescence-like quenching behavior, which protects the photosynthetic apparatus under strong light, but reduces the probability of effective capture~\cite{ruban2016,ruan2023}. Therefore, the functional role of aggregation is expected to be non-monotonic : moderate aggregation may improve spectral overlap and transmission, while excessive aggregation may introduce a competitive loss mechanism. This trade-off is particularly important for PSI-inspired artificial light harvesting design, where simply increasing the number of chromophore does not necessarily improve performance.

In this work, we construct a multi-layer conical exciton network inspired by PSI to study how the antenna geometry affects the excitation transmission efficiency. The model consists of a reaction center located at the apex of the conical structure and several antenna layers arranged around it. The site energy decreases from the outer antenna to the reaction center, forming a downhill energy funnel. The structural parameter $x$ controls the number of pigments in each layer, thereby adjusting the antenna density while maintaining the overall layered structure and energy gradient. The electron coupling is described by dipole-dipole interaction, and the excitation dynamics is simulated by combining the quantum jump approach proposed by~\citet{Piilo2008PRL} with the coherent modified Redfield theory (QJA-CMRT). This framework enables us to deal with coherent exciton evolution, environment-induced relaxation and dephasing, fluorescence loss, and irreversible reaction center trapping in the same dynamic model. The goal of this paper is to clarify why there is an optimal size for PSI-like antennas and how to understand this optimal value from both energy and dynamics perspectives. We show that increasing the pigment density enhances the intralayer coupling and broadens the exciton band, while the interlayer coupling remains almost unchanged. Therefore, the overlap of the density of states between adjacent layers increases first and then decreases, resulting in a non-monotonic dependence of the transmission efficiency on the antenna density. The population dynamics further reveals that the optimal network accelerates the transmission to the reaction center and reduces the fluorescence loss, while excessive delocalization at a large density weakens the effective directionality of the energy funnel. These results suggest that efficient PSI-like excitation energy transfer requires balanced spectral matching, exciton delocalization and directional trapping, rather than simply maximizing the number or coupling strength of pigments.

\section{Model and Methods}
\subsection{PSI-Inspired Quantum Network}
\label{sec:network}

We consider a quantum network model inspired by the structural organization and 
energy transfer mechanisms of natural Photosystem I (PSI), which is known for its 
highly efficient excitation transport~\cite{jordan2001,romero2017quantum,fromme2001,nelson2004}. In our model, 
the network consists of antenna pigments and a reaction center embedded within a 
finite three-dimensional region, with geometric parameters guided by the PSI 
architecture~\cite{sener2007}.

\begin{figure}[htbp]
\includegraphics[width=13.9 cm]{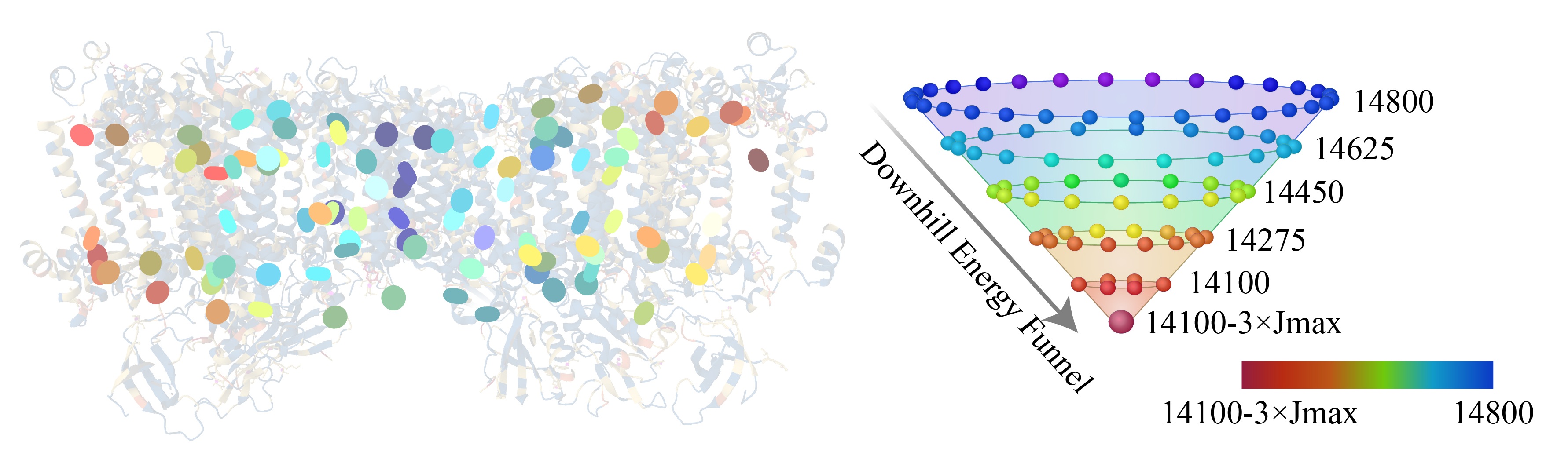}
\caption{A schematic diagram of the multi-layer conical quantum network model inspired by PSI. The left side is a natural PSI antenna structure~\cite{jordan2001,pdb1jb0}, and the right side is a three-dimensional conical structure. The antenna pigments are layered and arranged around the central reaction center (RC). The color gradient represents the site energy that forms the downhill energy funnel, and the energy value of each layer (in cm$^{-1}$) has been marked. }
\label{fig:schematic_diagram}
\end{figure}  
\unskip

As shown in figure~\ref{fig:schematic_diagram}, the antenna pigments are arranged in discrete layers surrounding the reaction center, forming an elongated conical architecture reminiscent of the PSI core 
antenna~\cite{jordan2001,benShem2003,Toporik2019}. Within each layer, pigments are distributed along an 
elliptical contour, with the layer radius increasing proportionally with the 
distance from the reaction center. This conical stacking maintains an 
approximately uniform pigment density throughout the network. The reaction 
center is located at the apex of the cone. 
The spatial dimensions of the network, namely $X_{\mathrm{len}}=124.23 $ \AA, $Y_{\mathrm{len}}=31.97$ \AA~
and $Z_{\mathrm{len}}=94.29$ \AA, are taken from the crystallographic data of the PSI core antenna~\cite{jordan2001} to ensure that the model geometry is consistent with the natural system.
The site energies of pigments are assigned such that they decrease progressively 
from the outermost antenna layer toward the reaction center, forming an effective 
energy funnel that guides excitation transport to the trap. The reaction center 
itself is positioned at a slightly lower energy than the innermost antenna layer, 
ensuring that the final capture step is energetically downhill and therefore 
irreversible. Such a combination of spatial organization and energetic gradients 
has been identified as a key design principle in natural photosynthetic systems~\cite{vangrondelle2006,romero2017quantum,renger2011modeling,ai2013clustered}.

The excitonic Hamiltonian of the network is written as
\begin{equation}
H = \sum_{n} \epsilon_n \ket{n}\bra{n} + \sum_{m \neq n} J_{mn} \ket{m}\bra{n},
\end{equation}
where $\epsilon_n$ denotes the site energy of pigment $n$, and $J_{mn}$ 
represents the electronic coupling between pigments $m$ and $n$~\cite{frenkel1931,mayKuhn2011}.
Electronic couplings are modeled using the dipole--dipole interaction~\cite{forster1948}
\begin{equation}
J_{mn} = 5042\, D^2
\left[
\frac{\bm{d}_m \cdot \bm{d}_n}{r_{mn}^3}
-3\frac{(\bm{d}_m \cdot \bm{r}_{mn})(\bm{d}_n \cdot \bm{r}_{mn})}{r_{mn}^5}
\right],
\end{equation}
where $\bm{d}_m$ and $\bm{d}_n$ denote the unit vectors of the transition dipole 
moments of pigments $m$ and $n$, $\bm{r}_{mn}$ is the displacement vector between 
them, and $D = 4.48$ Debye is the magnitude of the transition dipole moment 
(typical for chlorophyll $a$)~\cite{forster1948}. Distances are expressed in \AA.
For simplicity, all transition dipoles are assumed to be parallel and oriented 
along a fixed direction. This idealized assumption removes orientational disorder 
and allows us to isolate the influence of spatial geometry on excitation energy 
transport.

To reduce computational complexity, weak electronic couplings below a threshold 
value $J_c = 1\,\mathrm{cm^{-1}}$ are neglected~\cite{renger2001,scholak2011optimal}. Such truncation schemes are 
commonly employed in excitonic network models~\cite{scholak2011optimal,montepietra2020modeling}. 
The Hamiltonian matrix is constructed to be Hermitian.
The resulting model provides a simplified yet physically motivated framework 
that captures the essential structural features of PSI while enabling systematic 
exploration of how network architecture influences excitation energy transfer 
efficiency.

\subsection{Quantum Jump Dynamics with Loss Channels}
\label{sec:qj_dynamics}

To describe excitation energy transfer (EET) in the PSI-inspired network, 
we employ a quantum jump approach~\cite{Piilo2008PRL,Piilo2009PRA} (QJA) combined with the coherent modified Redfield theory (CMRT), 
following Ref.~\cite{li2025qja,ai2014qja}. 
This approach maps the evolution of the reduced density matrix onto an ensemble of stochastic 
wave-function trajectories, allowing an efficient treatment of large systems 
while retaining the influence of environmental interactions~\cite{plenio1998quantum,breuer2002,weiss2012}.
We first diagonalize the system Hamiltonian in the site basis to obtain the exciton representation. 
In this basis, the system Hamiltonian takes the form
\begin{equation}
H_{\mathrm{s}} = \sum_{\alpha} E_{\alpha} \ket{\alpha}\bra{\alpha},
\end{equation}
where $\ket{\alpha}$ denotes the exciton eigenstate and $E_{\alpha}$ is its corresponding energy. 
The interaction with the environment induces both population relaxation and pure dephasing~\cite{mayKuhn2011}, 
which are described by CMRT through a set of transition rates $R_{\alpha\beta}(t)$ and 
dephasing rates $\Gamma_{\alpha}(t)$~\cite{ishizaki2009,rebentrost2009,hwang-fu2015}.
To incorporate irreversible processes, the system is extended by including auxiliary states 
representing fluorescence loss and reaction center trapping. Within this framework, 
fluorescence corresponds to decay from exciton states to a loss channel with rate $R_{\mathrm{fl}}$, 
while the reaction center acts as a trapping channel with rate $R_{\mathrm{cr}}$. 
These processes introduce additional decay pathways that compete with coherent energy transfer~\cite{cao2009,sener2007}.

A distinctive feature of this approach is the treatment of non-Markovian effects~\cite{Piilo2008PRL,Piilo2009PRA}. 
The rates obtained from CMRT may become temporarily negative, reflecting environmental memory. 
Following the QJA formalism, transition channels are separated into positive ($j_+$, with $R_{j_+}(t) > 0$) 
and negative ($j_-$, with $R_{j_-}(t) < 0$) contributions~\cite{rebentrost2009nmqj}. 
The deterministic evolution of each trajectory is then governed by an effective non-Hermitian Hamiltonian 
that explicitly includes both types of channels,
\begin{equation}
H = H_{\mathrm{s}} - \frac{i}{2}\sum_{j_+} R_{j_+}(t) A_{j_+}^\dagger A_{j_+} - \frac{i}{2}\sum_{j_-} R_{j_-}(t) A_{j_-}^\dagger A_{j_-},
\label{eq:nonherm}
\end{equation}
where $A_{j_\pm}$ denote the jump operators associated with each channel~\cite{li2025qja}. 
The wave function evolves continuously as
\begin{equation}
\ket{\psi(t+dt)} = \frac{e^{-i H dt} \ket{\psi(t)}}{\| e^{-i H dt} \ket{\psi(t)} \|}.
\end{equation}
At each time step, this deterministic evolution is interrupted by quantum jumps. 
For a positive channel, the jump probability is given by
\begin{equation}
P_{j_+}(t) = \braket{\psi(t) | A_{j_+}^\dagger A_{j_+} | \psi(t)} \times R_{j_+}(t)\, dt,
\end{equation}
and the state after the jump becomes $A_{j_+} \ket{\psi(t)} / \| A_{j_+} \ket{\psi(t)} \|$. 
Positive channels account for conventional population transfer between exciton states, 
fluorescence emission, and charge separation at the reaction center~\cite{dalibard1992,plenio1998quantum}. Negative channels, 
on the other hand, induce reversed transitions that partially restore coherence, 
and their jump rules involve the population distribution across the ensemble~\cite{rebentrost2009nmqj}. 
This mechanism ensures that the stochastic evolution faithfully reproduces 
the underlying master equation even beyond the Markovian regime.

The physical density matrix is reconstructed by averaging over a large number of independent trajectories,
\begin{equation}
\rho(t) = \frac{1}{M} \sum_{k=1}^{M} \ket{\psi_k(t)}\bra{\psi_k(t)},
\end{equation}
where $M$ is the number of realizations. 
The initial excitation is placed on the antenna pigments, 
and the system is propagated until a steady state is reached.
The overall quantum efficiency is defined as the population accumulated in the reaction center 
in the long-time limit,
\begin{equation}
\eta = P_{\mathrm{RC}}(t \to \infty),
\end{equation}
which represents the probability that an excitation is successfully trapped~\cite{cao2009}. 
This formulation provides a unified and efficient framework to capture the interplay 
between coherent dynamics, environmental effects, 
and irreversible processes in large photosynthetic networks.
\section{Numerical Simulation}
\label{sec:simulation}

The numerical calculations are performed based on the PSI-inspired excitonic 
network introduced in Sec.~\ref{sec:network}. For each network configuration, 
the system Hamiltonian is constructed according to the prescribed pigment 
geometry, site energies, and dipole--dipole couplings. Long-range interactions 
with magnitudes below the cutoff $J_c = 1\,\mathrm{cm^{-1}}$ are neglected, 
yielding a sparse Hamiltonian that preserves the dominant energy transfer 
pathways while reducing computational overhead.

The size of the network is controlled by a single structural parameter $x$, 
which determines the number of pigments in the innermost antenna layer and consequently 
sets the total number of antenna sites $N_H$ as described in Sec.~\ref{sec:network}. 
In the present study, $x$ is varied over a suitable range to explore the 
influence of network density on energy transfer performance, while all other 
geometric and energetic parameters are held fixed at their PSI-inspired values.
For each network realization, the energy transfer efficiency $\eta$ is computed 
using the quantum jump framework detailed in Sec.~\ref{sec:qj_dynamics}. 
Numerical convergence with respect to the propagation time step and the maximum 
evolution time is carefully monitored. The steady-state condition is deemed to 
be reached when the trapping probability at the reaction center exhibits no 
appreciable drift over an extended interval, ensuring that the extracted 
efficiency $\eta$ is well converged.

\section{Results and Discussion}
\label{sec:results}

\begin{figure}[htbp]
\includegraphics[width=9 cm]{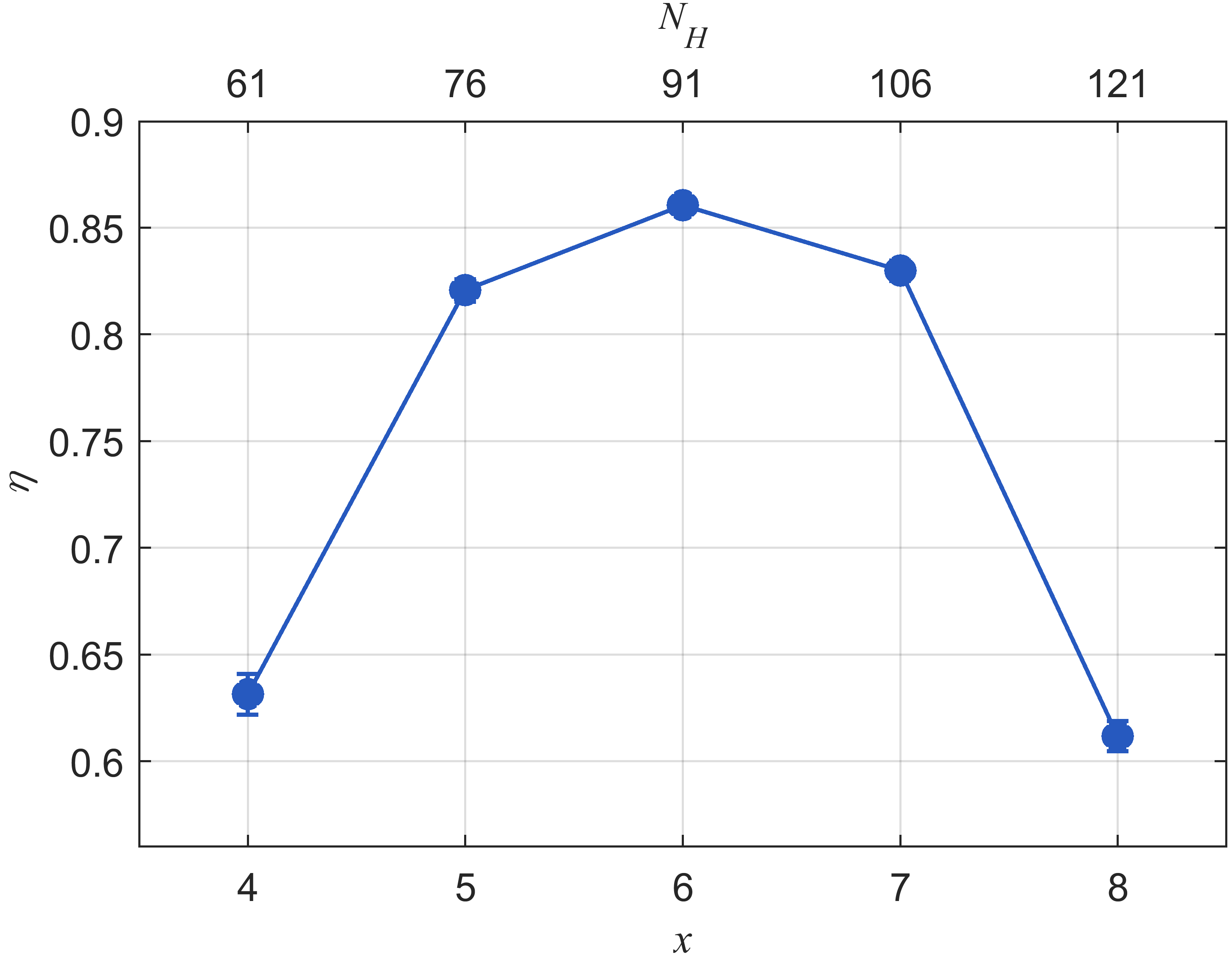}
\caption{Excitation energy transfer efficiency $\eta$ as a function of $x$ for $m=5$. 
The error bars represent the standard deviations over different initial excitation sites. 
The maximum efficiency $\eta=0.86$ is obtained at $x=6$ ($N_H=91$).}
\label{fig:eta_vs_x_m5}
\end{figure}  
\unskip

Since the overall transfer efficiency reaches its maximum at $m=5$, 
we use this configuration as the reference for detailed analysis. 
In this case, the average inter-layer energy gap is fixed at $\Delta E=700/(m-1)=175~\mathrm{cm^{-1}}$, 
and only the structural parameter $x$ is varied. 
Here, $x$ denotes the number of pigments in the innermost antenna layer adjacent to the reaction center, 
while the $n$th layer contains $nx$ pigments.
Figure~\ref{fig:eta_vs_x_m5} shows that the transfer efficiency exhibits 
a pronounced non-monotonic dependence on $x$. 
The efficiency increases from $\eta=0.63$ at $x=4$ to a maximum value of $\eta=0.86$ at $x=6$, 
and then decreases to $\eta=0.61$ at $x=8$. Although the total number of pigments, 
\begin{equation}
    N_H=1+\sum_{n=1}^{m}nx=1+15x,
\end{equation}
increases from 61 to 121, the transfer efficiency does not improve 
monotonically with system size. Instead, an optimal antenna size emerges, 
indicating that increasing the number of pigments alone is insufficient to enhance excitation transport.

To understand the origin of this non-monotonic behavior, 
we first examine the exciton band structures within individual antenna layers. 
Each layer is treated as a one-dimensional ring of strongly coupled pigments. 
The exciton bandwidths are obtained by diagonalizing the corresponding intra-layer sub-Hamiltonians. 
Table~\ref{tab:coupling_data} summarizes the maximum intra-layer coupling, 
the maximum inter-layer coupling, and the exciton bandwidths for different values of $x$.

\begin{table}[htbp]
\centering
\begin{threeparttable}
\small
\setlength{\tabcolsep}{12pt}
\caption{Intra- and inter-layer coupling strengths and exciton bandwidths for $m=5$.}
\label{tab:coupling_data}
\begin{tabular}{c c c c c c c c}
\toprule
\multirow{2}{*}{$x$} & 
\multirow{2}{*}{\shortstack{$|J_1|$ \\ (cm$^{-1}$)}} & 
\multirow{2}{*}{\shortstack{$|J_2|$ \\ (cm$^{-1}$)}} & 
\multicolumn{5}{c}{Exciton bandwidths (cm$^{-1}$)} \\
\cmidrule(lr){4-8}
 & & & $W_{L1}$ & $W_{L2}$ & $W_{L3}$ & $W_{L4}$ & $W_{L5}$ \\
\midrule
4 & 26.7 & 13.8 & 107.0 & 92.4 & 83.7 & 81.7 & 80.0 \\
5 & 54.2 & 13.8 & 192.6 & 174.2 & 163.9 & 162.8 & 161.2 \\
6 & 88.7 & 13.8 & 333.4 & 295.1 & 283.0 & 279.8 & 278.5 \\
7 & 128.7 & 13.8 & 493.6 & 464.3 & 448.0 & 444.8 & 441.9 \\
8 & 178.8 & 13.8 & 739.0 & 686.9 & 669.6 & 665.4 & 663.4 \\
\bottomrule
\end{tabular}
\begin{tablenotes}
\item[1] $x$ is the number of sites in the first layer. $J_1$ and $J_2$ are the maximum of intra- and inter-layer couplings. 
$W_{Ln}$ represents the widths of the energy band of the $n$th layer, which are obtained by diagonalizing the intra-layer sub-Hamiltonian of the $n$th layer.
\end{tablenotes}
\end{threeparttable}
\end{table}

As shown in Table~\ref{tab:coupling_data}, the intra-layer coupling increases monotonically with $x$, 
rising from $26.7~\mathrm{cm^{-1}}$ at $x=4$ to $178.8~\mathrm{cm^{-1}}$ at $x=8$. 
This trend originates from the increased pigment density within each layer, 
which reduces the average intermolecular distance and strengthens the dipole-dipole interaction. 
In contrast, the inter-layer coupling remains nearly constant at approximately $13.8~\mathrm{cm^{-1}}$.
The exciton bandwidths broaden significantly with increasing $x$. 
For example, the bandwidth of the first layer increases from $107.0~\mathrm{cm^{-1}}$ to $739.0~\mathrm{cm^{-1}}$. 
Moreover, the inner layers consistently possess larger bandwidths than the outer layers, 
reflecting their stronger couplings and larger numbers of pigments.
Because the intra-layer coupling is much stronger than the inter-layer coupling when $x\geq4$, the network exhibits a hierarchical structure consisting of strongly coupled layers connected 
by relatively weak inter-layer interactions. Exciton bands are therefore formed primarily 
within individual layers~\cite{knox1963,vanAmerongen2000}, and energy transfer between neighboring layers is largely determined by 
the spectral overlap between their exciton bands~\cite{forster1948,scholes2003,jang2004,mirkovic2017}.

\begin{figure}[htbp]
\includegraphics[width=9 cm]{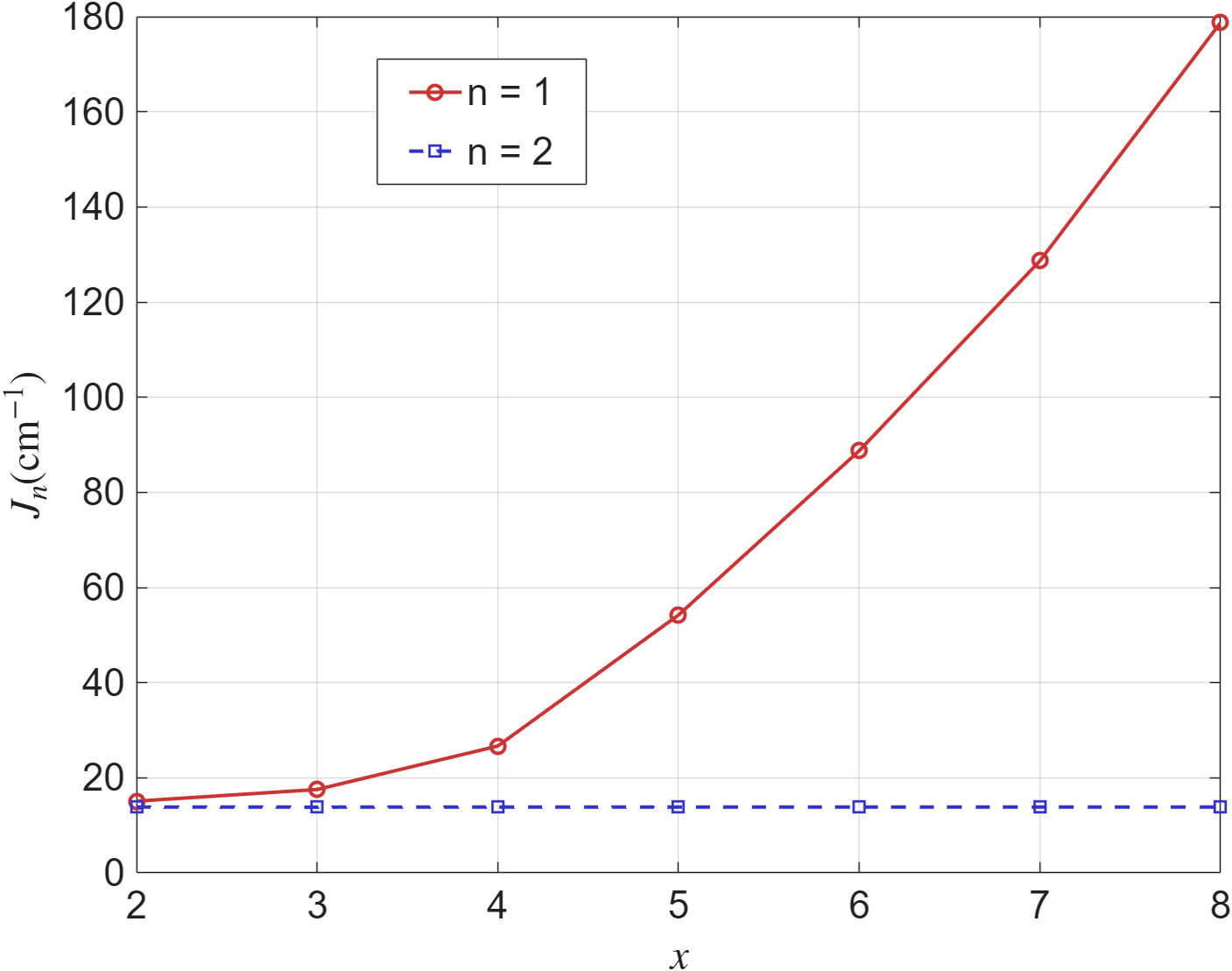}
\caption{Maximum intra-layer and inter-layer couplings $J_n$ (n=1, 2) as functions of $x$ for $m=5$. 
It can be seen that $J_1$ increases with the increase of $x$, and $J_2$ is independent of $x$. 
At $x\geqslant 5$, $J_1$ is much larger than $J_2$.}
\label{fig:coupling}
\end{figure}
\unskip  

To quantify spectral matching between neighboring layers, we define the
excitonic density of states of layer~$n$ by Gaussian-broadening its
eigenenergies,
\begin{equation}
D_n(E) = \frac{1}{N_n}\sum_{k=1}^{N_n}
\frac{1}{\sqrt{2\pi}\,\sigma}
\exp\!\left[
-\frac{\left(E - E_k^{(n)}\right)^2}{2\sigma^2}
\right],
\label{eq:DOS}
\end{equation}
where $E_k^{(n)}$ is the $k$-th eigenenergy of layer $n$, $N_n$ the
number of excitonic states, and $\sigma$ the broadening arising from
static disorder and environmental fluctuations~\cite{malyshev2001,scholes2003,mayKuhn2011}. 
The $1/N_n$ prefactor
normalizes $D_n$ to unit area, ensuring that the overlap reflects
genuine spectral similarity rather than state-count scaling. The
Gaussian form is motivated by the central limit theorem for many
independent perturbations and, unlike a Lorentzian, ensures exponential
suppression of overlap between well-separated bands~\cite{dahlbom2001,mukamel1995}. 
The pairwise and total spectral overlaps are defined as
\begin{equation}
S_{n,n+1}
= \int_{-\infty}^{\infty}
D_n(E)\,D_{n+1}(E)\,dE,
\qquad
S = \sum_{n=1}^{m-1} S_{n,n+1},
\label{eq:overlap}
\end{equation}
which evaluate analytically to
\begin{equation}
S_{n,n+1}
= \frac{1}{N_n N_{n+1}}\,
\frac{1}{2\sqrt{\pi}\,\sigma}
\sum_{k=1}^{N_n}\sum_{l=1}^{N_{n+1}}
\exp\!\left[
-\frac{\left(E_k^{(n)} - E_l^{(n+1)}\right)^2}{4\sigma^2}
\right].
\label{eq:analytic}
\end{equation}
Larger $S$ indicates stronger resonant matching and thus faster interlayer transfer. 
This index generalizes the spectral-overlap factor governing F\"orster and multichromophoric resonance energy transfer rates~\cite{jang2004,jang2007,scholes2003,renger2001}; 
since the inter-layer coupling ($\sim 14~\mathrm{cm}^{-1}$) is nearly
$x$-independent, variations in the transfer efficiency $\eta$ are
expected to be dominated by variations of 
$S$.
We set $\sigma = 10~\mathrm{cm}^{-1}$, consistent with the chlorophyll
\textit{a} linewidth in protein environments at cryogenic to
intermediate temperatures~\cite{chenu2015,vanAmerongen2000,spano2010}.
This value is comparable to the inter-layer coupling yet an order of
magnitude below the inter-layer energy gap
($\Delta E = 175~\mathrm{cm}^{-1}$), capturing the relevant transfer
scale without erasing layer-level spectral distinction.

\begin{figure}[htbp]
\includegraphics[width=9 cm]{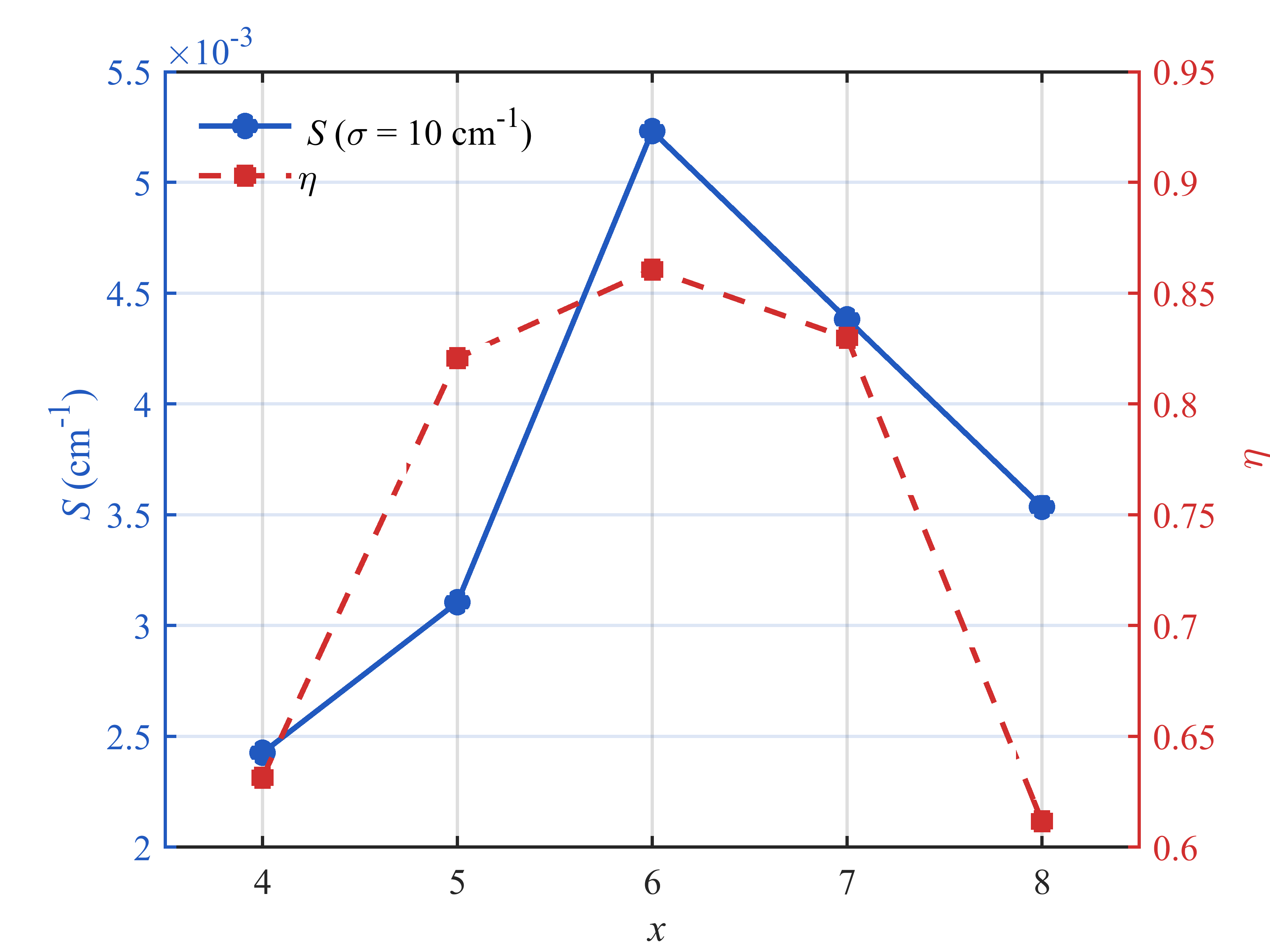}
\caption{Total density-of-states overlap $S$ (blue solid curve, left axis) 
and transfer efficiency $\eta$ (red dashed curve, right axis) as functions of 
antenna density $x$ for $m=5$ ($\sigma = 10~\mathrm{cm}^{-1}$). 
The two curves display a pronounced non-monotonic dependence on $x$ and 
reach their maxima at $x=6$ ($S \approx 5.23\times 10^{-3}~\mathrm{cm}^{-1}$, 
$\eta \approx 0.86$), indicating that spectral matching between neighboring 
exciton bands is a determinant of the transfer efficiency.}
\label{fig:linear_corr}
\end{figure}
\unskip

The total overlap $S$ and transmission efficiency $\eta$ show obvious non-monotonic dependence with the change of $x$, and reach the maximum value in $x=6$ ( $S = 5.23\times 10^{- 3}~\mathrm{cm}^{- 1}$, $\eta = 0.86$ ).
The close correspondence between the two curves confirms that spectral matching 
between adjacent layers is a key determinant of excitation transport. 
At $x=4$, the bandwidth of the innermost layer is smaller than 
$\Delta E$, limiting interlayer overlap. At $x=8$, the bandwidth 
broadens substantially, reducing the effective funnel directionality and 
weakening spectral selectivity between neighboring layers despite the 
increased number of states. The optimal configuration at $x=6$ thus 
balances these competing effects.
The close correspondence between $S(x)$ and $\eta(x)$ in
Figure~\ref{fig:linear_corr} confirms that spectral overlap is a
dominant factor governing interlayer transfer. As $x$ increases from 4 to 6,
both quantities increase monotonically, indicating that the broadening of the exciton bands
enhances the spectral overlap between adjacent layers, which in turn
facilitates resonant interlayer transport and increases the transfer
efficiency. Beyond $x=6$, the continued broadening of the bands weakens
the effective funnel directionality. Although the total number of
resonant states remains large, the loss of spectral selectivity between
neighboring layers causes both $S$ and $\eta$ to decrease. This
qualitative agreement indicates that spectral matching is a primary
driver of the non-monotonic efficiency, while other factors such as
exciton delocalization and energy-funnel directionality modulate the
efficiency beyond what spectral overlap alone captures.
To assess the robustness of these results, we varied $\sigma$ over the
range $5 \sim 50~\mathrm{cm}^{-1}$. The non-monotonic trend $S(x)$ and 
the synchronous variation of $S$ and $\eta$ with $x$ are preserved
throughout this range, confirming that the conclusions are insensitive to
the precise choice of the broadening parameter.

\begin{figure}[htbp]
\includegraphics[width=13.9 cm]{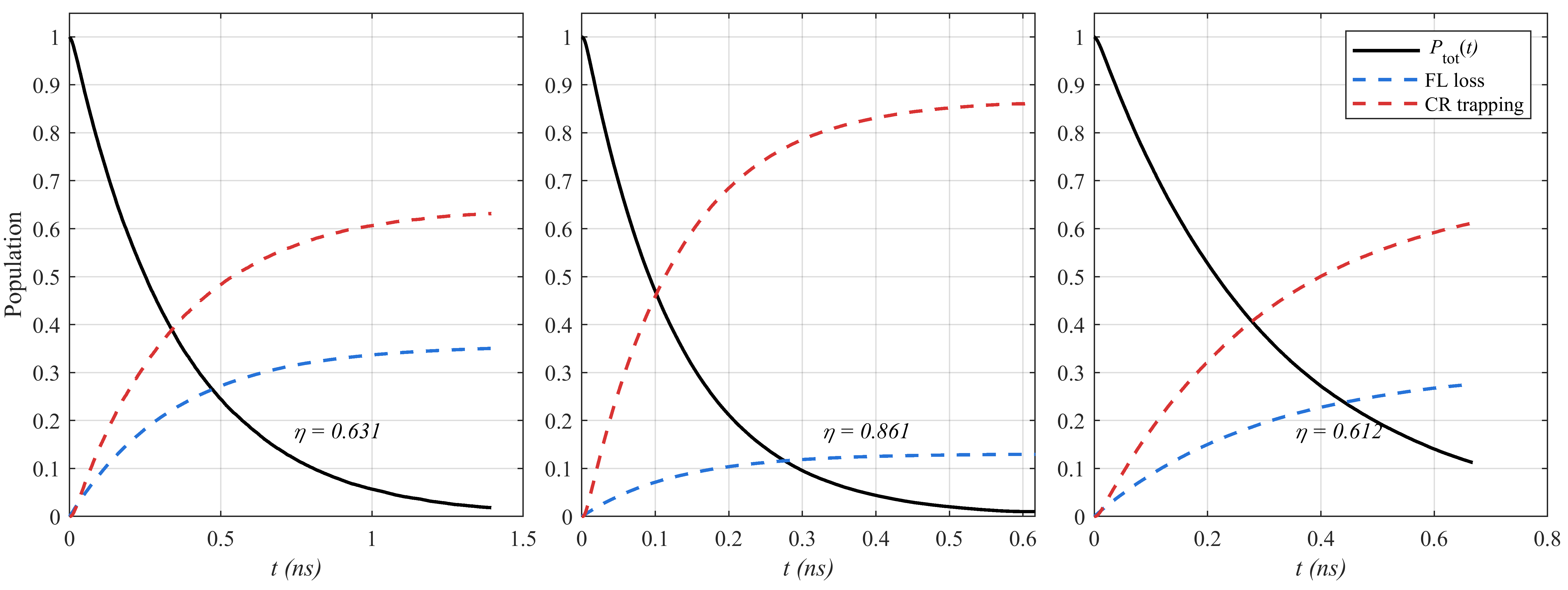}
\caption{Ensemble-averaged population dynamics for the five-layer network at three representative antenna densities: (left) $x=4$ ($\eta=0.63$), 
(center) $x=6$ ($\eta=0.86$), and (right) $x=8$ ($\eta=0.61$).  
The inset in each panel displays the short-time population relaxation $P_{\rm rel}(t)$ 
within the first 600~fs, reflecting the initial exciton thermalization process. 
At $x=6$, the total population decays most rapidly and the CR channel captures 
the largest fraction of excitations, whereas at $x=4$ and $x=8$ a substantial 
fraction is diverted to the FL loss channel.}
\label{fig:kinetics}
\end{figure}
\unskip

The influence of spectral overlap on transport is directly reflected in the population dynamics. 
Figure~\ref{fig:kinetics} presents the ensemble-averaged population evolution together 
with the fluorescence (FL) and reaction-center (CR) channels. The characteristic decay time, 
estimated using the $1/e$ criterion, decreases from approximately $3.6\times10^{5}$ fs
at $x=4$ to $1.3\times10^{5}$ fs
at $x=6$, and then increases again to about $3.1\times10^{5}$ fs
at $x=8$.
The competition between the CR and FL channels determines the final efficiency. 
At $x=6$, the CR channel dominates and most excitations are successfully trapped by the reaction center, 
resulting in the highest efficiency. In contrast, larger fractions of excitation are lost 
through fluorescence at $x=4$ and $x=8$.
The population dynamics are consistent with the picture inferred from the exciton band structures. 
At small $x$, insufficient spectral overlap slows down interlayer transfer and increases fluorescence loss. 
Increasing $x$ enhances the overlap and accelerates transport toward the reaction center, 
leading to the optimal efficiency at $x=6$. Further increasing $x$ reduces the total density-of-states 
overlap and causes the efficiency to decrease.

\begin{figure}[htbp]
\includegraphics[width=13.9 cm]{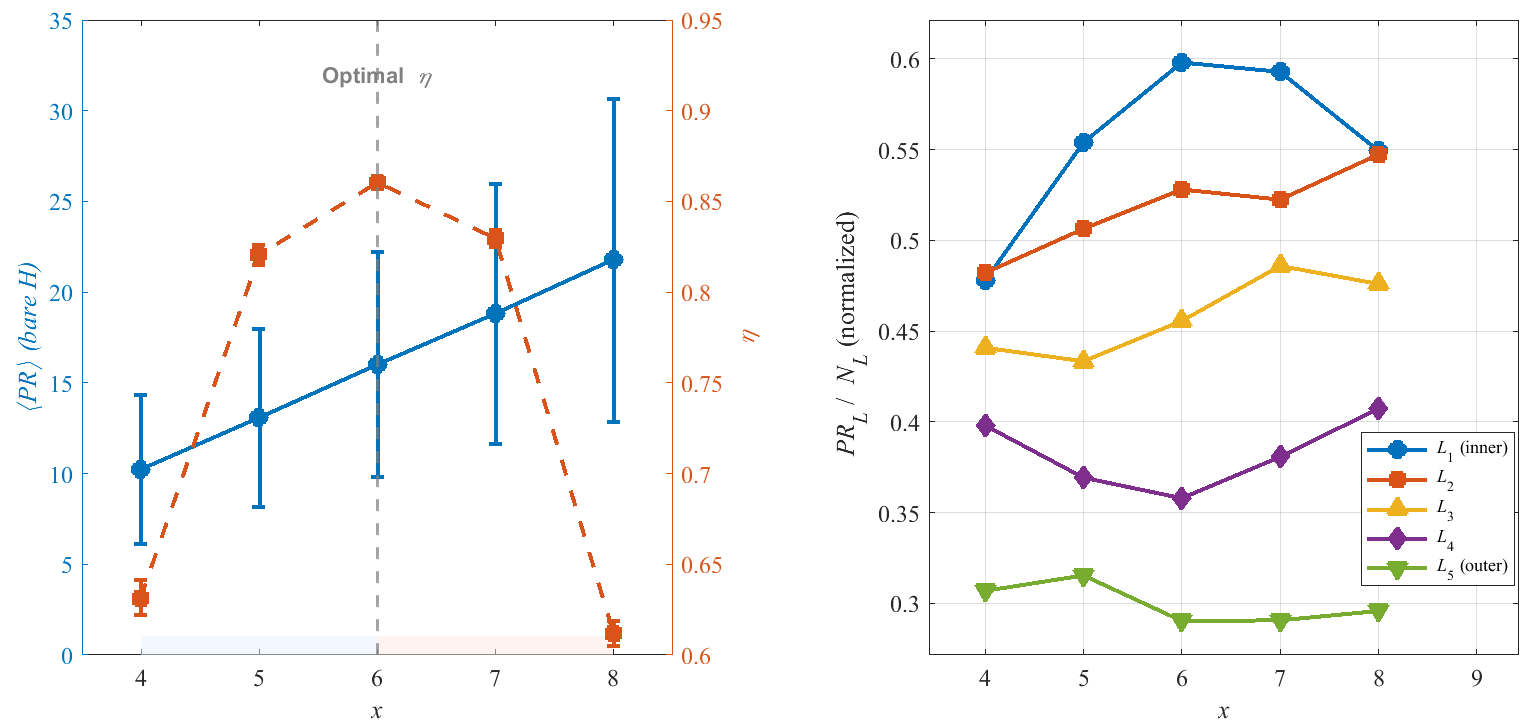}
\caption{The left side shows that the ensemble-averaged participation ratio $\langle \mathrm{PR} \rangle$ of the bare 
exciton Hamiltonian (blue, left axis) and transfer efficiency $\eta$ (red, right axis) 
as functions of $x$ for $m=5$. The error bars on $\langle \mathrm{PR} \rangle$ represent 
the standard deviation of PR values across all $N_H$ exciton eigenstates, 
$\sigma_{\mathrm{PR}} = \sqrt{\frac{1}{N_H}\sum_\alpha (\mathrm{PR}_\alpha - \langle \mathrm{PR} \rangle)^2}$, 
quantifying the spread in the degree of localization among different eigenstates. 
Both $\langle \mathrm{PR} \rangle$ and $\sigma_{\mathrm{PR}}$ increase monotonically with $x$, 
reflecting both stronger delocalization and greater heterogeneity of exciton states. 
The efficiency $\eta$ reaches its maximum at $x=6$ and then declines, 
indicating a trade-off between exciton mobility and funnel directionality. 
The right side is the layer-resolved normalized participation ratio $\langle \mathrm{PR} \rangle_i/N_{\mathrm{L}_i}$ 
for the $i$th antenna layer. The normalized participation ratio increases most strongly 
in the outer antenna layers, indicating that exciton states become progressively more 
extended across the network as $x$ increases.}
\label{fig:ipr_combined}
\end{figure}
\unskip

To understand why fluorescence loss becomes more pronounced at larger $x$, 
we examine the exciton participation ratio (PR), which quantifies the degree of 
delocalization of exciton eigenstates over the pigment network. 
The PR of the $\alpha$-th exciton eigenstate is defined as $\mathrm{PR}_\alpha = 1/\sum_n |c_{\alpha n}|^4$, 
where $c_{\alpha n} = \langle n|\alpha\rangle$ is the amplitude on site $n$~\cite{scholes2011,chenu2015,dahlbom2001,spano2010}. 
Physically, PR approximates the effective number of pigments over which the exciton 
wavefunction is significantly distributed---a fully localized state gives $\mathrm{PR} = 1$, 
while a state uniformly delocalized over all $N_H$ sites gives $\mathrm{PR} = N_H$. 
The corresponding inverse participation ratio is $\mathrm{IPR}_\alpha = \sum_n |c_{\alpha n}|^4$, 
which quantifies the degree of exciton localization~\cite{malyshev2001}. 
A larger PR indicates that the exciton wavefunction is spread over a greater number of pigments~\cite{fassioli2014}.
The left side of Figure~\ref{fig:ipr_combined} shows the ensemble-averaged PR alongside the efficiency. 
The PR increases monotonically with $x$, from approximately 10 at $x=4$ to approximately 22 at $x=8$, 
reflecting the progressive delocalization of exciton states as the intra-layer coupling strengthens. 
The error bars represent the standard deviation of PR values across all $N_H$ eigenstates; 
their increase with $x$ indicates that the exciton states become not only more delocalized on average, 
but also more heterogeneous---some states remain relatively localized while others extend across large 
portions of the network. 
For $x\le 6$, both the DOS overlap and the transfer efficiency increase with $x$, 
indicating that spectral matching is the dominant factor governing transport. 
For $x > 6$, however, the efficiency decreases despite the continued growth of 
the participation ratio. This observation suggests that exciton delocalization alone 
does not determine the transport efficiency~\cite{huelga2013,mohseni2008}. 
Instead, the optimal performance emerges when a large density-of-states overlap 
is achieved without excessively reducing the effective funnel directionality~\cite{huelga2013}.
To resolve where in the network the additional delocalization occurs, we perform a 
layer-resolved analysis of the participation ratio. For each exciton eigenstate, 
we compute its PR within each layer after renormalizing the amplitude restricted to 
that layer, $\mathrm{PR}_{\alpha}^{(L)} = 1/\sum_{n\in L} |\tilde{c}_{\alpha n}|^4$, 
and then average over all eigenstates and normalize by the number of sites $N_{\mathrm{L}_i}$ 
in layer $i$ to enable cross-layer comparison. 
The right side of Figure~\ref{fig:ipr_combined} shows that the normalized participation ratio increases 
most noticeably in the outer layers. This trend indicates that the exciton eigenstates 
become progressively more extended across the antenna network as $x$ increases, 
consistent with the stronger intra-layer couplings and broader exciton bands. 
The increased delocalization reduces the spatial distinction between neighboring layers 
and weakens the effective directionality imposed by the energy funnel, impairing efficient 
inward transport toward the reaction center.

In summary, the optimal efficiency at $m=5$ and $x=6$ originates primarily from the maximization of 
the density-of-states overlap between neighboring exciton bands. Increasing $x$ strengthens 
the intra-layer coupling and broadens the exciton bands, thereby enhancing spectral overlap 
and facilitating interlayer transport. The total overlap reaches its maximum at $x=6$, 
giving rise to the highest transfer efficiency.
For larger $x$, the density-of-states overlap decreases despite the continued increase 
in exciton delocalization. The resulting reduction in spectral matching, together 
with the weakening of the effective funnel directionality associated with increasingly extended 
exciton states~\cite{lloyd2011}, leads to a gradual decline in efficiency.
Therefore, the optimal antenna size emerges from a balance between efficient spectral matching 
and the preservation of directional energy transport toward the reaction center. 
{
This theoretical prediction concurs with recent experimental findings in coccolithophore PSI,
where a layered arrangement of 35-38 LHC antennae sustains $\sim$95\% quantum efficiency
despite a markedly expanded absorption cross-section \cite{shen2025science,sun2025sciadv}.
Physiologically, this modular expansion achieved through concentric peripheral antenna layers
rather than increased core pigment density reflects a key evolutionary adaptation in
photosynthetic lineages \cite{Zhang2024natplants,Naschberger2024natplants}.
It exemplifies the optimal balance between spectral broadening and directional energy funneling,
validating our model's premise that geometric architecture is central to efficient light
harvesting \cite{kato2024elife,Feng2024jipb}.}

\section{Conclusions}
\label{sec:conclusions}

In this work, we systematically investigated how the geometric architecture of 
a PSI-inspired conical antenna network influences excitation energy transfer (EET). 
Using the quantum jump approach combined with the coherent modified Redfield theory (QJA-CMRT), 
we studied a series of layered excitonic networks in which the antenna density was controlled by 
the structural parameter $x$, while the interlayer energy gradient was fixed at 
$\Delta E = 175~\mathrm{cm^{-1}}$ for a five-layer ($m=5$) structure.

The simulations reveal a pronounced non-monotonic dependence of the transfer efficiency on network size. 
As $x$ increases, the efficiency rises from $\eta=0.63$ at $x=4$ to a maximum value of $\eta=0.86$ at $x=6$, 
corresponding to a network containing $N_H=91$ pigments. Further increasing the network size leads to 
a gradual decrease in efficiency. These results demonstrate that adding more pigments does not 
necessarily improve transport performance and that an optimal antenna size naturally emerges 
from the interplay between network geometry and excitonic dynamics.
Analysis of the exciton band structure shows that increasing $x$ strengthens 
the intra-layer coupling and substantially broadens the exciton bandwidths, 
whereas the interlayer coupling remains nearly unchanged. As a result, energy transfer 
between neighboring layers is primarily controlled by the spectral overlap between their exciton bands. 
By explicitly evaluating the density-of-states overlap, we find that the total overlap increases 
from $x=4$ to $x=6$ and reaches its maximum at the optimal configuration. The close correspondence 
between the overlap integral and the transfer efficiency indicates 
that spectral matching between adjacent layers is a key factor governing excitation transport.
The population dynamics further support this interpretation. 
Networks with larger density-of-states overlap exhibit faster population relaxation 
and more efficient trapping by the reaction center. In contrast, reduced overlap leads to 
slower interlayer transport and increased fluorescence loss. The shortest characteristic 
decay time and the strongest reaction-center trapping are both observed at $x=6$, 
consistent with the maximum transfer efficiency.
To understand the efficiency reduction at larger antenna sizes, we analyzed 
the exciton participation ratio (PR), which characterizes the spatial extent of exciton states. 
The average PR increases monotonically with $x$, indicating progressively stronger exciton delocalization. 
Moderate delocalization is beneficial because it facilitates exciton migration across the antenna network. 
However, when the exciton states become excessively extended, the effective directionality imposed by the energy funnel is weakened. 
In this regime, excitations are distributed over a larger fraction of the network and become more susceptible to fluorescence loss 
before reaching the reaction center. The decrease in efficiency for $x>6$ therefore originates 
from the combined effects of reduced spectral matching and excessive exciton delocalization.

Overall, the optimal transport efficiency arises from a balance between two competing mechanisms. 
Increasing network density enhances the density-of-states overlap between neighboring layers 
and promotes interlayer transport, whereas excessive bandwidth broadening leads to stronger 
exciton delocalization and reduced funnel directionality. The optimal configuration occurs 
when these two effects are simultaneously balanced, yielding the highest trapping efficiency.
Interestingly, the optimal network identified here ($x=6$, $N_H=91$) contains a number of 
pigments comparable to that of the natural PSI core antenna~\cite{benShem2003,Toporik2019}, 
which consists of approximately 90--100 chlorophyll \textit{a} molecules~\cite{jordan2001}. 
This agreement suggests that the interplay between spectral overlap and exciton delocalization may represent a general design principle underlying the organization of natural photosynthetic systems. From an evolutionary perspective, the fact that the optimal antenna size coincides with the core antenna size of cyanobacterial PSI~\cite{karapetyan1999}, the most ancient oxygenic photosystem, suggests that this optimal pigment density may have been fixed early in the evolution of photosynthesis. Subsequent expansion of the light-harvesting capacity across different lineages was achieved by adding modular peripheral LHC subunits in a layered arrangement~\cite{nelson2015,Croce2013}, a strategy that preserves the optimal pigment density within each layer while extending the overall absorption cross-section, consistent with the design principle identified in our theoretical analysis.
More broadly, the present work demonstrates that the QJA-CMRT framework provides an efficient 
and physically transparent approach for investigating large-scale excitonic networks. 
The observed optimization mechanism emerges naturally from geometric and energetic constraints, 
without invoking additional quantum-enhancement assumptions. These findings provide useful insights 
for the design of artificial light-harvesting systems, where tuning network geometry offers 
a practical route toward optimizing excitation energy transport.

\section{Abbreviations}

The following abbreviations are used in this manuscript:
\\

\noindent 
\begin{tabular}{@{}ll}
EET & Excitation energy transfer\\
PSI & Photosystem I\\
QJA & Quantum jump approach\\
CMRT & Coherence modified Redfield theory\\
RC & Reaction center\\
FL & Fluorescence\\
PR & Participation ratio
\end{tabular}

\section{Acknowledgments}
This work is supported by Guangdong Provincial Quantum Science Strategic Initiative under Grant no. GDZX2505004, 
and Innovation Program for Quantum Science and Technology under Grant no. 2023ZD0300200, 
the National Natural Science Foundation of China under Grant no. 62461160263.

\bibliography{photosynthesis_refs_processed}

@article{Croce2013,
  author  = {Croce, Roberta and van Amerongen, Herbert},
  title   = {Light-Harvesting in Photosystem {I}},
  journal = {Photosynth. Res.},
  year    = {2013},
  volume  = {116},
  pages   = {153--166},
  doi     = {10.1007/s11120-013-9838-x}
}

@article{nelson2015,
  author  = {Nelson, Nathan and Junge, Wolfgang},
  title   = {Structure and Energy Transfer in Photosystems of Oxygenic Photosynthesis},
  journal = {Annu. Rev. Biochem.},
  year    = {2015},
  volume  = {84},
  pages   = {659--683},
  doi     = {10.1146/annurev-biochem-092914-041942}
}

@article{akhtar2021,
  author  = {Akhtar, Parveen and Caspy, Ido and Nowakowski, Pawe{\l} J. and Malavath, Tirupathi and Nelson, Nathan and Tan, Howe-Siang and Lambrev, Petar H.},
  title   = {Two-Dimensional Electronic Spectroscopy of a Minimal Photosystem {I} Complex Reveals the Rate of Primary Charge Separation},
  journal = {J. Am. Chem. Soc.},
  year    = {2021},
  volume  = {143},
  pages   = {14601--14612},
  doi     = {10.1021/jacs.1c05010}
}

@article{renger2009,
  author  = {Renger, Thomas},
  title   = {Theory of Excitation Energy Transfer: From Structure to Function},
  journal = {Photosynth. Res.},
  year    = {2009},
  volume  = {102},
  pages   = {471--485},
  doi     = {10.1007/s11120-009-9472-9}
}

@article{mattioni2021,
  author  = {Mattioni, Andrea and Caycedo-Soler, Felipe and Huelga, Susana F. and Plenio, Martin B.},
  title   = {Design Principles for Long-Range Energy Transfer at Room Temperature},
  journal = {Phys. Rev. X},
  year    = {2021},
  volume  = {11},
  pages   = {041003},
  doi     = {10.1103/PhysRevX.11.041003}
}

@article{ruan2023,
  author  = {Ruan, Meixia and Li, Hao and Zhang, Ying and Zhao, Ruoqi and Zhang, Jun and Wang, Yingjie and Gao, Jiali and Wang, Zhuan and Wang, Yumei and Sun, Dapeng and Ding, Wei and Weng, Yuxiang},
  title   = {Cryo-{EM} Structures of {LHCII} in Photo-Active and Photo-Protecting States Reveal Allosteric Regulation of Light Harvesting and Excess Energy Dissipation},
  journal = {Nat. Plants},
  year    = {2023},
  volume  = {9},
  pages   = {1547--1557},
  doi     = {10.1038/s41477-023-01500-2}
}

@article{ruban2016,
  author  = {Ruban, Alexander V.},
  title   = {Nonphotochemical Chlorophyll Fluorescence Quenching: Mechanism and Effectiveness in Protecting Plants from Photodamage},
  journal = {Plant Physiol.},
  year    = {2016},
  volume  = {170},
  pages   = {1903--1916},
  doi     = {10.1104/pp.15.01935}
}

@article{romero2017quantum,
  title     = {Quantum Design of Photosynthesis for Bio-Inspired Solar-Energy Conversion},
  author    = {Romero, Elisabet and Novoderezhkin, Vladimir I. and van Grondelle, Rienk},
  journal   = {Nature},
  volume    = {543},
  number    = {7645},
  pages     = {355--365},
  year      = {2017},
  publisher = {Nature Publishing Group},
  doi       = {10.1038/nature22012}
}

@article{ishizaki2009,
  author    = {Ishizaki, Akihito and Fleming, Graham R.},
  title     = {Unified Treatment of Quantum Coherent and Incoherent Hopping Dynamics in Electronic Energy Transfer: Reduced Hierarchy Equation Approach},
  journal   = {J. Chem. Phys.},
  volume    = {130},
  number    = {23},
  pages     = {234111},
  year      = {2009},
  doi       = {10.1063/1.3155372}
}

@article{chin2013,
  author    = {Chin, Alex W. and Prior, Javier and Rosenbach, R. and Caycedo-Soler, Felipe and Huelga, Susana F. and Plenio, Martin B.},
  title     = {The Role of Non-Equilibrium Vibrational Structures in Electronic Coherence and Recoherence in Pigment--Protein Complexes},
  journal   = {Nat. Phys.},
  volume    = {9},
  number    = {2},
  pages     = {113--118},
  year      = {2013},
  doi       = {10.1038/nphys2515}
}

@article{plenio2008,
  author    = {Plenio, Martin B. and Huelga, Susana F.},
  title     = {Dephasing-Assisted Transport: Quantum Networks and Biomolecules},
  journal   = {New J. Phys.},
  volume    = {10},
  number    = {11},
  pages     = {113019},
  year      = {2008},
  doi       = {10.1088/1367-2630/10/11/113019}
}

@article{mohseni2008,
  author    = {Mohseni, Masoud and Rebentrost, Patrick and Lloyd, Seth and Aspuru-Guzik, Al{\'a}n},
  title     = {Environment-Assisted Quantum Walks in Photosynthetic Energy Transfer},
  journal   = {J. Chem. Phys.},
  volume    = {129},
  number    = {17},
  pages     = {174106},
  year      = {2008},
  doi       = {10.1063/1.3002335}
}

@article{rebentrost2009,
  author    = {Rebentrost, Patrick and Mohseni, Masoud and Kassal, Ivan and Lloyd, Seth and Aspuru-Guzik, Al{\'a}n},
  title     = {Environment-Assisted Quantum Transport},
  journal   = {New J. Phys.},
  volume    = {11},
  number    = {3},
  pages     = {033003},
  year      = {2009},
  doi       = {10.1088/1367-2630/11/3/033003}
}

@article{jordan2001,
  author    = {Jordan, Patrick and Fromme, Petra and Witt, Horst Tobias and Klukas, Olaf and Saenger, Wolfram and Krau{\ss}, Norbert},
  title     = {Three-Dimensional Structure of Cyanobacterial Photosystem {I} at 2.5 {\AA} Resolution},
  journal   = {Nature},
  volume    = {411},
  number    = {6840},
  pages     = {909--917},
  year      = {2001},
  doi       = {10.1038/35082000}
}

@article{scholes2011,
  author    = {Scholes, Gregory D. and Fleming, Graham R. and Olaya-Castro, Alexandra and van Grondelle, Rienk},
  title     = {Lessons from Nature About Solar Light Harvesting},
  journal   = {Nat. Chem.},
  volume    = {3},
  number    = {10},
  pages     = {763--774},
  year      = {2011},
  doi       = {10.1038/nchem.1145}
}

@article{vangrondelle2006,
  author    = {van Grondelle, Rienk and Novoderezhkin, Vladimir I.},
  title     = {Energy Transfer in Photosynthesis: Experimental Insights and Quantitative Models},
  journal   = {Phys. Chem. Chem. Phys.},
  volume    = {8},
  number    = {7},
  pages     = {793--807},
  year      = {2006},
  doi       = {10.1039/B514032C}
}

@article{ai2013clustered,
  author    = {Ai, Qing and Yen, Tzu-Chi and Jin, Bih-Yaw and Cheng, Yuan-Chung},
  title     = {Clustered Geometries Exploiting Quantum Coherence Effects for Efficient Energy Transfer in Light Harvesting},
  journal   = {J. Phys. Chem. Lett.},
  volume    = {4},
  number    = {15},
  pages     = {2577--2584},
  year      = {2013},
  doi       = {10.1021/jz4011477}
}

@article{ai2014qja,
  author    = {Ai, Qing and Fan, Yuan-Jia and Jin, Bih-Yaw and Cheng, Yuan-Chung},
  title     = {An Efficient Quantum Jump Method for Coherent Energy Transfer Dynamics in Photosynthetic Systems under the Influence of Laser Fields},
  journal   = {New J. Phys.},
  volume    = {16},
  pages     = {053033},
  year      = {2014},
  doi       = {10.1088/1367-2630/16/5/053033}
}

@article{li2025qja,
  author    = {Li, Rui and Li, Yiteng and Zhang, Ke-Ying and Ai, Qing},
  title     = {Quantum Jump Approach for Photosynthetic Energy Transfer with Chemical Reaction and Fluorescence Loss},
  journal   = {Innov. Energy},
  volume    = {2},
  pages     = {100114},
  year      = {2025},
  doi       = {10.59717/j.xinn-energy.2025.100114}
}

@article{cao2009,
  author    = {Cao, Jianshu and Silbey, Robert J.},
  title     = {Optimization of Exciton Trapping in Energy Transfer Processes},
  journal   = {J. Phys. Chem. A},
  volume    = {113},
  number    = {50},
  pages     = {13825--13838},
  year      = {2009},
  doi       = {10.1021/jp9032589}
}

@article{sener2007,
  author    = {{\c{S}}ener, Melih K. and Schulten, Klaus and Park, Sanghyun},
  title     = {F{\"o}rster Energy Transfer Theory as Reflected in the Structures of Photosynthetic Light-Harvesting Systems},
  journal   = {Chem. Phys.},
  volume    = {341},
  number    = {1-3},
  pages     = {9--17},
  year      = {2007},
  doi       = {10.1016/j.chemphys.2007.04.010}
}

@article{scholak2011optimal,
  author    = {Scholak, Torsten and Wellens, Thomas and Buchleitner, Andreas},
  title     = {Optimal Networks for Excitonic Energy Transport},
  journal   = {J. Phys. B: At. Mol. Opt. Phys.},
  volume    = {44},
  number    = {18},
  pages     = {184012},
  year      = {2011},
  doi       = {10.1088/0953-4075/44/18/184012}
}

@article{plenio1998quantum,
  author    = {Plenio, Martin B. and Knight, Peter L.},
  title     = {The Quantum-Jump Approach to Dissipative Dynamics in Quantum Optics},
  journal   = {Rev. Mod. Phys.},
  volume    = {70},
  number    = {1},
  pages     = {101--144},
  year      = {1998},
  doi       = {10.1103/RevModPhys.70.101}
}

@article{rebentrost2009nmqj,
  author    = {Rebentrost, Patrick and Chakraborty, Rupak and Aspuru-Guzik, Al{\'a}n},
  title     = {Non-Markovian Quantum Jumps in Excitonic Energy Transfer},
  journal   = {J. Chem. Phys.},
  volume    = {131},
  number    = {18},
  pages     = {184102},
  year      = {2009},
  doi       = {10.1063/1.3259838}
}

@article{renger2011modeling,
  author    = {Renger, Thomas},
  title     = {Modeling of Photosynthetic Light-Harvesting: From Structure to Function},
  journal   = {Procedia Chem.},
  volume    = {3},
  pages     = {236--247},
  year      = {2011},
  doi       = {10.1016/j.proche.2011.08.031}
}

@article{montepietra2020modeling,
  author = {Montepietra, D. and Bellingeri, M. and Ross, A. M. and Scotognella, F. and Cassi, D.},
    title = {Modelling photosystem I as a complex interacting network},
    journal = {J. R. Soc. Interface},
    volume = {17},
    number = {172},
    pages = {20200813},
    year = {2020},
    issn = {1742-5689},
    doi = {10.1098/rsif.2020.0813}
}

@article{mirkovic2017,
  author = {Mirkovic, Tihana and Ostroumov, Evgeny E. and Anna, Jessica M. and van Grondelle, Rienk and Govindjee and Scholes, Gregory D.},
  title   = {Light Absorption and Energy Transfer in the Antenna Complexes of Photosynthetic Organisms},
  journal = {Chem. Rev.},
  volume  = {117},
  number  = {2},
  pages   = {249--293},
  year    = {2017},
  doi     = {10.1021/acs.chemrev.6b00002},
}

@article{scholes2003,
  author  = {Scholes, Gregory D.},
  title   = {Long-Range Resonance Energy Transfer in Molecular Systems},
  journal = {Annu. Rev. Phys. Chem.},
  volume  = {54},
  pages   = {57--87},
  year    = {2003},
  doi     = {10.1146/annurev.physchem.54.011002.103746},
}

@article{hwang-fu2015,
title = {A coherent modified Redfield theory for excitation energy transfer in molecular aggregates},
journal = {Chem. Phys.},
volume = {447},
pages = {46-53},
year = {2015},
issn = {0301-0104},
doi = {https://doi.org/10.1016/j.chemphys.2014.11.026},
url = {https://www.sciencedirect.com/science/article/pii/S0301010414003358},
author = {Yu-Hsien Hwang-Fu and Wei Chen and Yuan-Chung Cheng}
}

@article{huelga2013,
author = {S.F. Huelga and M.B. Plenio},
title = {Vibrations, quanta and biology},
journal = {Contemp. Phys.},
volume = {54},
number = {4},
pages = {181--207},
year = {2013},
publisher = {Taylor \& Francis},
doi = {10.1080/00405000.2013.829687}
}

@article{benShem2003,
title={Crystal Structure of Plant Photosystem I},
author={Ben-Shem, A. et al.},
journal={Nature},
volume={426},
pages={630--635},
year={2003},
doi = {10.1038/nature02200}
}

@article{forster1948,
title={Zwischenmolekulare Energiewanderung und Fluoreszenz},
author={F\"{o}rster, Theodor},
journal={Ann. Phys. (Berlin)},
volume={437},
pages={55--75},
year={1948},
doi = {10.1002/andp.19484370105}
}

@article{frenkel1931,
title={On the Transformation of Light into Heat in Solids},
author={Frenkel, J.},
journal={Phys. Rev.},
volume={37},
issue = {1},
pages={17--44},
year={1931},
doi = {10.1103/PhysRev.37.17}
}

@article{fassioli2014,
title={Photosynthetic Light Harvesting: Excitons and Coherence},
author={Fassioli, Francesco et al.},
journal={J. R. Soc. Interface},
volume={11},
number={92},
pages={20130901},
year={2014},
issn = {1742-5689},
doi = {10.1098/rsif.2013.0901}
}

@article{lloyd2011,
title={Quantum Coherence in Biological Systems},
author={Lloyd, Seth},
journal={J. Phys.: Conf. Ser.},
volume={302},
pages={012037},
year={2011},
doi={10.1088/1742-6596/302/1/012037}
}

@article{knox1963,
  author  = {Knox, Ronald S.},
  title   = {Theory of {Excitons}},
  journal = {Solid State Phys.},
  volume  = {5},
  pages   = {Suppl.},
  year    = {1963},
  publisher = {Academic Press},
  address = {New York}
}

@article{fromme2001,
  author  = {Fromme, Petra and Jordan, Patrick and Krauss, Norbert},
  title   = {Structure of {Photosystem} {I}},
  journal = {Biochim. Biophys. Acta},
  volume  = {1507},
  number  = {1--2},
  pages   = {5--22},
  year    = {2001},
  doi     = {10.1016/S0005-2728(01)00199-9}
}

@article{nelson2004,
  author  = {Nelson, Nathan and Ben-Shem, Amit},
  title   = {The complex architecture of oxygenic photosynthesis},
  journal = {	Nat. Rev. Mol. Cell Biol.},
  volume  = {5},
  number  = {12},
  pages   = {971--982},
  year    = {2004},
  doi     = {10.1038/nrm1525}
}

@book{breuer2002,
  author  = {Breuer, Heinz-Peter and Petruccione, Francesco},
  title   = {The {Theory} of {Open} {Quantum} {Systems}},
  publisher = {Oxford University Press},
  address = {Oxford},
  year    = {2002},
  isbn    = {978-0-19-852063-4}
}

@book{weiss2012,
  author  = {Weiss, Ulrich},
  title   = {Quantum {Dissipative} {Systems}},
  edition = {4th},
  publisher = {World Scientific},
  address = {Singapore},
  year    = {2012},
  isbn    = {978-981-4374-91-8}
}

@article{chenu2015,
  author  = {Chenu, Aur{\'e}lia and Scholes, Gregory D.},
  title   = {Coherence in energy transfer and photosynthesis},
  journal = {Annu. Rev. Phys. Chem.},
  volume  = {66},
  pages   = {69--96},
  year    = {2015},
  doi     = {10.1146/annurev-physchem-040214-121713}
}

@article{renger2001,
  author  = {Renger, Thomas and May, Volkhard and K{\"u}hn, Oliver},
  title   = {Ultrafast excitation energy transfer dynamics in photosynthetic pigment--protein complexes},
  journal = {Phys. Rep.},
  volume  = {343},
  number = {3},
  pages   = {137--254},
  year    = {2001},
  issn = {0370-1573},
  doi     = {10.1016/S0370-1573(00)00078-8}
}

@article{dalibard1992,
  author  = {Dalibard, Jean and Castin, Yvan and M{\"o}lmer, Klaus},
  title   = {Wave-function approach to dissipative processes in quantum optics},
  journal = {Phys. Rev. Lett.},
  volume  = {68},
  number  = {5},
  pages   = {580--583},
  year    = {1992},
  doi     = {10.1103/PhysRevLett.68.580}
}

@inbook{may2011,
author = {Oliver, Kühn and Volkhard, May},
publisher = {John Wiley\&Sons, Ltd},
isbn = {9783527696277},
title = {Excitation Energy Transfer},
booktitle = {Charge and Energy Transfer Dynamics in Molecular Systems},
chapter = {9},
pages = {421-514},
doi = {https://doi.org/10.1002/9783527696277.ch9},
url = {https://onlinelibrary.wiley.com/doi/abs/10.1002/9783527696277.ch9},
eprint = {https://onlinelibrary.wiley.com/doi/pdf/10.1002/9783527696277.ch9},
year = {2023}
}

@article{jang2004,
  author  = {Jang, Seogjoo and Newton, Marshall D. and Silbey, Robert J.},
  title   = {Multichromophoric {F}\"orster Resonance Energy Transfer},
  journal = {Physi. Rev. Lett.},
  volume  = {92},
  pages   = {218301},
  year    = {2004},
  doi     = {10.1103/PhysRevLett.92.218301}
}

@article{malyshev2001,
  author  = {Malyshev, Victor A. and Malyshev, Andrei V.},
  title   = {Statistics of Low-Energy Excitons and Optical Line Shapes},
  journal = {Phys. Rev. B},
  volume  = {63},
  pages   = {195111},
  year    = {2001},
  doi     = {10.1103/PhysRevB.63.195111}
}

@article{spano2010,
  author  = {Spano, Frank C.},
  title = {The Spectral Signatures of Frenkel Polarons in H- and J-Aggregates},
  journal = {Acc. Chem. Res.},
  volume  = {43},
  pages   = {429--439},
  year    = {2010},
  doi     = {10.1021/ar900233v}
}

@book{vanAmerongen2000,
  author    = {van Amerongen, Herbert and Valkunas, Leonas
               and van Grondelle, Rienk},
  title     = {Photosynthetic Excitons},
  publisher = {World Scientific},
  address   = {Singapore},
  year      = {2000},
  isbn      = {978-981-02-3280-1}
}

@book{mayKuhn2011,
  author    = {May, Volkhard and K\"uhn, Oliver},
  title     = {Charge and Energy Transfer Dynamics in Molecular Systems},
  edition   = {3},
  publisher = {Wiley-VCH},
  address   = {Weinheim},
  year      = {2011},
  isbn      = {978-3527407320}
}

@inproceedings{mukamel1995,
  title={Principles of Nonlinear Optical Spectroscopy},
  author    = {Mukamel, Shaul},
  year={1995},
  url={https://api.semanticscholar.org/CorpusID:118565806}
}

@article{dahlbom2001,
  author  = {Dahlbom, M. and Pullerits, T. and Mukamel, S. and Sundstr\"{o}m, V.},
  title   = {Exciton Delocalization in the {B850} Light-Harvesting Complex},
  journal = {J. Phys. Chem. B},
  volume  = {105},
  pages   = {5515--5524},
  year    = {2001},
  doi     = {10.1021/jp004496i}
}

@article{jang2007,
  author  = {Jang, Seogjoo and Newton, Marshall D. and Silbey, Robert J.},
  title   = {Theory of Multichromophoric Resonance Energy Transfer},
  journal = {J. Chem. Phys.},
  volume  = {127},
  pages   = {174710},
  year    = {2007},
  doi     = {10.1063/1.2800326}
}

@article{zhu2024quantum,
  author  = {Zhu, Rui and Li, Wei and Zhen, Zhe and Zou, Jun and Liao, Guang and Wang, Jie and Wang, Zhuan and Chen, Hong and Qin, Song and Weng, Yuxiang},
  title   = {Quantum Phase Synchronization via Exciton--Vibrational Energy Dissipation Sustains Long-Lived Coherence in Photosynthetic Antennas},
  journal = {Nat. Commun.},
  year    = {2024},
  volume  = {15},
  pages   = {3171},
  doi     = {10.1038/s41467-024-47560-6}
}

@article{wang2024incoherent,
  author  = {Wang, Jie and Zhu, Rui and Zou, Jun and Liu, Hao and Meng, Hao and Zhen, Zhe and Li, Wei and Wang, Zhuan and Chen, Hong and Pu, Ying and Weng, Yuxiang},
  title   = {Incoherent Ultrafast Energy Transfer in Phycocyanin 620 from 	extit{Thermosynechococcus vulcanus} Revealed by Polarization-Controlled Two-Dimensional Electronic Spectroscopy},
  journal = {J. Chem. Phys.},
  year    = {2024},
  volume  = {161},
  pages   = {085101},
  doi     = {10.1063/5.0222587}
}

@article{huchzermeyer1986direct,
  author  = {Huchzermeyer, Bernhard and Loehr, A. and Willms, I.},
  title   = {A Direct Interaction between Photosystem {I} and the Chloroplast Coupling Factor},
  journal = {Biochem. J.},
  year    = {1986},
  volume  = {234},
  number  = {1},
  pages   = {217--220},
  doi     = {10.1042/bj2340217}
}

@article{huchzermeyer1981coupling,
  author  = {Huchzermeyer, Bernhard},
  title   = {The Coupling of Electron Flow to {ATP} Synthesis in Pea and Maize Mesophyll Chloroplasts: {I.} {Interaction} of Adenine Nucleotides and Energy Transfer Inhibitors with the Coupling Factor Complex},
  journal = {Plant Physiology},
  volume  = {68},
  number  = {3},
  pages   = {610--615},
  year    = {1981},
  doi     = {10.1104/pp.68.3.610},
}

@article{huchzermeyer1980bicarbonate,
  author  = {Huchzermeyer, Bernhard},
  title   = {A Proposed Mechanism for the Stimulatory Effect of Bicarbonate Ions on {ATP} Synthesis in Isolated Chloroplasts},
  journal = {Plant Physiology},
  volume  = {66},
  number  = {2},
  pages   = {242--245},
  year    = {1980},
  doi     = {10.1104/pp.66.2.242},
}

@inbook{papenbrock2016current,
  author    = {Papenbrock, J. and Churakova, Y. and Huchzermeyer, B. and Koyro, H. W.},
  title     = {Current Methods in Photosynthesis Research},
  booktitle = {Handbook of Photosynthesis},
  edition   = {3},
  publisher = {CRC Press},
  year      = {2016},
  pages     = {447--464}
}

@article{song2021excitonic,
  author  = {Song, Yin and Sechrist, R. and Nguyen, H. H. and Johnson, W. and Abramavicius, D. and Redding, K. E. and Ogilvie, J. P.},
  title   = {Excitonic Structure and Charge Separation in the Heliobacterial Reaction Center Probed by Multispectral Multidimensional Spectroscopy},
  journal = {Nat. Commun.},
  year    = {2021},
  volume  = {12},
  pages   = {2801},
  doi     = {10.1038/s41467-021-23060-9}
}

@article{nguyen2023charge,
  author  = {Nguyen, H. H. and Song, Yin and Maret, E. L. and Silori, Y. and Willow, R. and Yocum, C. F. and Ogilvie, J. P.},
  title   = {Charge Separation in the Photosystem {II} Reaction Center Resolved by Multispectral Two-Dimensional Electronic Spectroscopy},
  journal = {Sci. Adv.},
  year    = {2023},
  volume  = {9},
  number  = {18},
  pages   = {eade7190},
  doi     = {10.1126/sciadv.ade7190}
}

@article{song2019vibronic,
  author  = {Song, Yin and Schubert, A. and Maret, E. and Burdick, R. K. and Dunietz, B. D. and Geva, E. and Ogilvie, J. P.},
  title   = {Vibronic Structure of Photosynthetic Pigments Probed by Polarized Two-Dimensional Electronic Spectroscopy and Ab Initio Calculations},
  journal = {Chem. Sci.},
  year    = {2019},
  volume  = {10},
  pages   = {8143--8153},
  doi     = {10.1039/C9SC02329A}
}

@article{Maly2019chemphys,
title = {Electron-vibrational coupling decreases trapping by low-energy states in photosynthesis},
author = {P. Malý and V.-I. Novoderezhkin and R. {van Grondelle} and T. Mančal},
journal = {Chem. Phys.},
volume = {522},
pages = {69-76},
year = {2019},
issn = {0301-0104},
doi = {https://doi.org/10.1016/j.chemphys.2019.02.011},
url = {https://www.sciencedirect.com/science/article/pii/S0301010418311133}
}

@article{Toporik2019,
  author    = {Toporik, Hila and Li, Jin and Williams, Dewight and Chiu, Po-Lin and Mazor, Yuval},
  title     = {The structure of the stress-induced photosystem {I}--{IsiA} antenna supercomplex},
  journal   = {	Nat. Struct. Mol. Biol.},
  year      = {2019},
  volume    = {26},
  number    = {6},
  pages     = {443--449},
  doi       = {10.1038/s41594-019-0228-8}
}

@misc{pdb1jb0,
  author       = {{RCSB PDB}},
  title        = {Crystal Structure of {Photosystem I}: a {Photosynthetic} {Reaction Center} and {Core Antenna System} from {Cyanobacteria}},
  howpublished = {Protein Data Bank},
  year         = {2001},
  doi          = {10.2210/pdb1JB0/pdb},
  url          = {https://www.rcsb.org/structure/1JB0}
}

@article{Piilo2008PRL,
  title = {{Non-Markovian} Quantum Jumps},
  author = {Piilo, J. and Maniscalco, S. and H\"ark\"onen, K. and Suominen, K.-A.},
  journal = {Phys. Rev. Lett.},
  volume = {100},
  issue = {18},
  pages = {180402},
  numpages = {4},
  year = {2008},
  month = {May},
  publisher = {American Physical Society},
  doi = {10.1103/PhysRevLett.100.180402},
  url = {https://link.aps.org/doi/10.1103/PhysRevLett.100.180402}
}

@article{Piilo2009PRA,
  title = {{Open system dynamics with non-Markovian quantum jumps}},
  author = {Piilo, J. and H\"ark\"onen, K. and Maniscalco, S. and Suominen, K.-A.},
  journal = {Phys. Rev. A},
  volume = {79},
  issue = {6},
  pages = {062112},
  numpages = {17},
  year = {2009},
  month = {06},
  publisher = {American Physical Society},
  doi = {https://journals.aps.org/pra/abstract/10.1103/PhysRevA.79.062112},
  url = {https://link.aps.org/doi/10.1103/PhysRevA.79.062112}
}

@article{Cao2020sciadv,
Author = {Cao, J.-S. and Cogdell, R.-J. and Coker, David F. and Duan,
   H.-G. and Hauer, J. and Kleinekathoefer, U. and Jansen,
   T.-L.-C. and Mancal, T. and Miller, R.-J. Dwayne and Ogilvie,
   J.-P. and Prokhorenko, I, V. and Renger, T. and Tan,
   H.-S. and Tempelaar, R. and Thorwart, M. and Thyrhaug,
   E. and Westenhoff, S. and Zigmantas, D.},
Title = {Quantum biology revisited},
Journal = {Sci. Adv.},
Year = {2020},
Volume = {6},
Number = {14},
Month = {APR},
DOI = {10.1126/sciadv.aaz4888},
url = {https://www.science.org/doi/10.1126/sciadv.aaz4888}
}

@book{Blankenship2014,
  author    = {Blankenship, Robert E.},
  title     = {Molecular Mechanisms of Photosynthesis},
  year      = {2014},
  edition   = {2},
  publisher = {Wiley-Blackwell},
  address   = {Oxford, UK},
  isbn      = {978-1-4051-8975-8},
  doi       = {10.1002/9781118658613},
  pages     = {72},
  chapter   = {5}
}

@article{Lambert2013nphys,
Author = {Lambert, N. and Chen, Y.-N. and Cheng, Y.-C. and Li, C.-M.
   and Chen, G.-Y. and Nori, F.},
Title = {Quantum biology},
Journal = {Nature Phys.},
Year = {2013},
Volume = {9},
Number = {1},
Pages = {10-18},
Month = {JAN},
DOI = {10.1038/NPHYS2474},
url = {https://www.nature.com/articles/nphys2474}
}

@article{Lee2007science,
Author = {Lee, H.-J. and Cheng, Y.-C. and Fleming, G.-R.},
Title = {Coherence dynamics in photosynthesis: Protein protection of excitonic
   coherence},
Journal = {Science},
Year = {2007},
Volume = {316},
Number = {5830},
Pages = {1462-1465},
Month = {JUN 8},
DOI = {10.1126/science.1142188},
url = {https://www.science.org/doi/epdf/10.1126/science.1142188}
}

@article{dong2012coherent,
  author  = {Dong, Hui and Xu, Da-Zhi and Huang, Jin-Feng and Sun, Chang-Pu},
  title   = {Coherent excitation transfer via the dark-state channel in a bionic system},
  journal = {Light: Sci. App.},
  volume  = {1},
  pages   = {e2},
  year    = {2012},
  doi     = {10.1038/lsa.2012.2},
}

@article{shen2025science,
  author  = {Shen, Lili and Ren, Fei and Wang, Yin-Chu and Li, Zhenhua and Zheng, Mengyuan and Li, Xiaoyi and Fan, Wenzheng and Yang, Yanyan and Sang, Min and Liu, Cheng and Han, Guangye and Qin, Song and Fan, Jianhua and Tian, Lijin and Kuang, Tingyun and Shen, Jian-Ren and Wang, Wenda},
  title   = {Structure and function of a huge photosystem {I}--fucoxanthin chlorophyll supercomplex from a coccolithophore},
  journal = {Science},
  year    = {2025},
  volume  = {389},
  number  = {6760},
  pages   = {eadv2132},
  doi     = {10.1126/science.adv2132},
}

@article{sun2025sciadv,
  author  = {Sun, Xiao-Meng and Li, Kang and He, Fei-Yu and Wang, Hao-Jie and Chen, Hai-Long and Wen, Quan and Liu, He-Yuan and Zhao, Fang and Chen, Xiu-Lan and Weng, Yu-Xiang and Gao, Jun and Liu, Lu-Ning and Zhang, Yu-Zhong and Zhao, Long-Sheng},
  title   = {Architecture and energy transfer of coccolithophore photosystem {I} with a huge light-harvesting antenna system},
  journal = {Science Advances},
  year    = {2025},
  volume  = {11},
  number  = {51},
  pages   = {eaea4965},
  doi     = {10.1126/sciadv.aea4965},
}

@article{kato2024elife,
  author = {Kato, Koji and Nakajima, Yoshiki and Xing, Jian and Kumazawa, Minoru and Ogawa, Haruya and Shen, Jian-Ren and Ifuku, Kentaro and Nagao, Ryo},
  title = {Structural basis for molecular assembly of fucoxanthin chlorophyll \textit{a}/\textit{c}-binding proteins in a diatom photosystem I supercomplex},
  journal = {eLife},
  year    = {2024},
  volume  = {13},
  pages = {RP99858},
  doi     = {10.7554/eLife.99858},
}

@article{Feng2024jipb,
  author  = {Feng, Yue and Li, Zhenhua and Yang, Yang and Shen, Lili and Li, Xiaoyi and Liu, Xueyang and Zhang, Xiaofei and Zhang, Jinyang and Ren, Fei and Wang, Yuan and Liu, Cheng and Han, Guangye and Wang, Xuchu and Kuang, Tingyun and Shen, Jian-Ren and Wang, Wenda},
  title   = {Structures of {PSI}--{FCPI} from \textit{Thalassiosira pseudonana} grown under high light provide evidence for convergent evolution and light-adaptive strategies in diatom {FCPIs}},
  journal = {Journal of Integrative Plant Biology},
  year    = {2024},
  volume  = {67},
  issue  = {4},
  pages   = {949--966},
  doi     = {10.1111/jipb.13816},
}

@article{Zhang2024natplants,
  author  = {Zhang, Aihong and Tian, Lin and Zhu, Tong and Li, Mengyu and Sun, Mengwei and Fang, Ying and Zhang, Yi and Lu, Congming},
  title   = {Uncovering the photosystem {I} assembly pathway in land plants},
  journal = {Nature Plants},
  year    = {2024},
  volume  = {10},
  pages   = {645--660},
  doi     = {10.1038/s41477-024-01658-3},
}

@article{Naschberger2024natplants,
  author  = {Naschberger, Andreas and Fadeeva, Mariia and Klaiman, Daniel and Borovikova-Sheinker, Anna and Caspy, Ido and Nelson, Nathan and Amunts, Alexey},
  title   = {Structure of plant photosystem {I} in a native assembly state defines {PsaF} as a regulatory checkpoint},
  journal = {Nature Plants},
  year    = {2024},
  volume  = {10},
  pages   = {874--879},
  doi     = {10.1038/s41477-024-01699-8},
}

@article{Tian2024ijms,
  author  = {Tian, Li-Rong and Chen, Jing-Hua},
  title   = {Photosystem {I}: A Paradigm for Understanding Biological Environmental Adaptation Mechanisms in Cyanobacteria and Algae},
  journal = {International Journal of Molecular Sciences},
  year    = {2024},
  volume  = {25},
  number  = {16},
  pages   = {8767},
  doi     = {10.3390/ijms25168767},
}

@article{Amelii2026natcomm,
  author  = {Amelii, Antonello and Capaldi, Stefano and Russo, Mattia and Guardini, Zeno and Sanit\'{a}, Gennaro and Esposito, Emanuela and Oliv\'{e}, Irene and Maiuri, Margherita and Dall'Osto, Luca and Cerullo, Giulio and Procaccini, Gabriele and Bassi, Roberto},
  title   = {Structural and spectral adaptation of the seagrass \textit{Posidonia oceanica} photosystem {I} to seabed light},
  journal = {Nature Communications},
  year    = {2026},
  volume  = {17},
  number  = {6135},
  doi     = {10.1038/s41467-026-75200-8},
}

@article{Akhtar2025cellrep,
  author  = {Akhtar, Parveen and Feng, Yue and Wang, Wenda and Shen, Jian-Ren and B\"{u}chel, Claudia and Tan, Howe-Siang and Lambrev, Petar H.},
  title = {Excitonic network organization enables efficient energy transfer in diatom photosystem I},
  journal = {Cell Reports Physical Science},
  year    = {2025},
  volume  = {6},
  number  = {8},
  pages = {102731},
  doi     = {10.1016/j.xcrp.2025.102731},
}

@article{karapetyan1999,
  author  = {Karapetyan, Navassard V. and Holzwarth, Alfred R. and R\"{o}gner, Matthias},
  title   = {The Photosystem {I} Trimer of Cyanobacteria: Molecular Organization, Excitation Dynamics and Physiological Significance},
  journal = {FEBS Letters},
  year    = {1999},
  volume  = {460},
  pages   = {395--400},
  doi     = {10.1016/S0014-5793(99)01352-6}
}

\end{document}